\documentclass[%
reprint,
superscriptaddress,
showpacs,preprintnumbers,
nofootinbib,
 amsmath,amssymb,
floatfix,
]{revtex4-1}
\usepackage[colorlinks=true,
urlcolor=blue,
linkcolor=darkraspberry,
citecolor=blue,
linktocpage=true,
pdfproducer=medialab,
pdfa=true,
anchorcolor=blue]{hyperref}%
\usepackage{xcolor}
\definecolor{cambridgeblue}{rgb}{0.64, 0.76, 0.68}
\definecolor{darkraspberry}{rgb}{0.53, 0.15, 0.34}
\usepackage{graphicx}
\usepackage{dcolumn}
\usepackage{bm}
\usepackage{hyperref}
\usepackage{amssymb}
\usepackage{placeins}

\usepackage{float}
\usepackage[utf8]{inputenc} 
\usepackage{xcolor}

\begin{document}

\title{Revisiting the scotogenic model with scalar dark matter}

\author{Ivania M. \'Avila}
\email{ivania.maturana@edu.uai.cl}
\affiliation{Departamento de Ciencias, Facultad de Artes Liberales, Universidad Adolfo Ib\'a\~{n}ez, Diagonal Las Torres 2640, Santiago, Chile}

\author{Giovanna Cottin}
\email{giovanna.cottin@uai.cl}
\affiliation{Departamento de Ciencias, Facultad de Artes Liberales, Universidad Adolfo Ib\'a\~{n}ez, Diagonal Las Torres 2640, Santiago, Chile}
\affiliation{Millennium Institute for Subatomic Physics at the High Energy Frontier (SAPHIR), \\ Fernández Concha 700, Santiago, Chile}

\author{Marco A. D\'iaz}
\email{mad@susy.fis.puc.cl}
\affiliation{Instituto de F\'isica, Pontificia Universidad Cat\'olica de Chile, Avenida Vicu\~{n}a Mackenna 4860, Santiago, Chile}
\affiliation{Millennium Institute for Subatomic Physics at the High Energy Frontier (SAPHIR), \\ Fernández Concha 700, Santiago, Chile}

\date{\today}


\begin{abstract}

The scotogenic model is a well motivated scenario that provides both an explanation for neutrino masses and for dark matter. We focus on a real scalar dark matter candidate in this model, produced through standard thermal freeze-out. We analyze the parameter space of the model compatible with the observed dark matter relic abundance,  direct and indirect detection searches, limits from lepton flavour violating decays and constraints from the neutrino sector. As the mass differences of the dark matter with the neutral and charged states are found to be small, the new scalars and fermions of the theory will have macroscopic lifetimes, and could thus be potentially detected with long-lived particle signatures at colliders. We find regions in the parameter space to be - partially or fully - consistent with the dark matter relic abundance, and the prediction of a long-lived charged scalar or lightest neutral fermion in the scotogenic scenario, for dark matter masses below 500 GeV. We discuss on the collider phenomenology in some detail. 

\end{abstract}

\maketitle

\section{Introduction}

Two of the main Standard Model (SM) of particle physics puzzles are the mechanism behind neutrino mass generation and the nature of dark matter (DM). Evidence on neutrino oscillations~\cite{Tanabashi:2018oca} and the observed energy density of the DM relic abundance~\cite{Aghanim:2018eyx} are clear indications that new physics is needed, which is ambitiously being looked for at experimental facilities worldwide, with yet null results. Testing simultaneously both hypothesis has inspired model building of several types, including the original scotogenic model~\cite{Ma:2006km} and extensions with additional phenomenological considerations (see for 
instance refs.~\cite{Hirsch:2013ola,Diaz:2016udz,Borah:2017dfn,Ahriche:2017iar,Hagedorn:2018spx,Ahriche:2018ger,Hugle:2018qbw,Avila:2019hhv,Leite:2020kos,Leite:2020bnb,Beniwal:2020hjc,DeRomeri:2021yjo,Barman:2021ifu}).

The original scotogenic model~\cite{Ma:2006km} is one of the simplest extensions of the SM that contains a DM candidate (which can be either bosonic or fermionic) and provides a viable explanation for small neutrino masses in the SM (via one-loop radiative corrections). A $\mathbb{Z}_{2}$ symmetry is imposed to forbid tree-level neutrino masses and to have a stable DM particle candidate. The new fields in the model consist of a scalar doublet, $\eta$, and three singlet fermionic fields $N_{i}$,  with $i = 1, 2, 3$. All these new fields are odd under the $\mathbb{Z}_{2}$ symmetry.

There has been recent interest in analyzing the collider signatures of the scotogenic model for different fermionic dark matter scenarios~\cite{Hessler:2016kwm,Hagedorn:2018spx,Baumholzer:2019twf,Baumholzer:2019twf}, and in models with similar low energy phenomenology~\cite{Borah:2018smz}. Particular attention in these works is given to novel, striking signatures at colliders, produced when the particles in the model are long-lived\footnote{For a review of long-lived particle searches at the Large Hadron Collider, see ref.~\cite{Alimena:2019zri}.}. In ref.~\cite{Hessler:2016kwm} the authors study a feebly interacting massive particle (FIMP) DM scenario, and investigate possible long-lived particle signatures at colliders depending on the mass spectrum of the $\mathbb{Z}_{2}$- odd particles. In ref.~\cite{Baumholzer:2019twf}, collider signatures were studied with a fermionic freeze-out DM, where the new scalars are heavier than the new fermions. 

In contrast, in this work we focus on a complementary scenario where the charged scalars of the new doublet ($\eta^{\pm}$) or the lightest neutral fermion ($N_{1}$) are long-lived particles, and the electroweak scale DM  (which we set as the real scalar component of $\eta$, $\eta_{R}$) is produced via standard thermal freeze-out, as envisioned in the original scotogenic model. An additional motivation to consider scalar dark matter is the possibility of comparison with the Inert Higgs Doublet model (IHDM)~\cite{Deshpande:1977rw}, where the particle content of the scalar sector is very similar to the scotogenic model, making the scalar phenomenology of both models hard to distinguish at particle colliders (see also~\cite{Maturana-Avila:2019qph}). Refs.~\cite{Borah:2017dfn,Sarma:2020msa} comment on a scalar DM candidate with an intermediate-mass region close to $\approx 500$ GeV produced via thermal freeze-out. This region is also called in the literature the IHDM {\it{desert}} region, where the dark matter relic density becomes under abundant~\cite{Borah:2017dfn,Barman:2021ifu} and direct detection searches constrains the allowed parameter space for $\mathcal{O}(100)$ GeV masses~\cite{Klasen:2013btp,Diaz:2015pyv}. The authors in~\cite{Borah:2017dfn,Sarma:2020msa} conclude that for fixed choices of couplings in the scalar potential and the lightest fermion $N_{1}$ of mass $m_{N_{1}}\geq 1$  TeV,  it is not possible to get the observed relic density below 500 GeV if the dark matter is produced only thermally in the scotogenic scenario. Nevertheless, in ref.~\cite{Klasen:2013jpa} it was first demonstrated that a region consistent with dark matter abundance can exist for dark matter masses below 500 GeV, as it was shown that coannihilations with the right-handed fermions can modify the viable region allowed for dark matter, leading to a GeV scale scalar dark matter candidate that satisfies the total relic density allowed by WMAP~\cite{WMAP:2010qai}.

In this work, we focus on the same region first identifyied in ref.~\cite{Klasen:2013jpa} -- for scalar dark matter masses below 500 GeV -- confirming its existance, as opposed to what is claimed in refs.~\cite{Borah:2017dfn,Sarma:2020msa}. We perform an updated an detailed numerical scan with constraints from lepton number violation, neutrino physics, direct and indirect searches, as well as the computation of the dark matter abundance consistent with limits set by Planck~\cite{Aghanim:2018eyx}. We find consistent results with ref.~\cite{Klasen:2013jpa}, when a smaller mass for $N_1$, close to $500$ GeV or below, a smaller mass splitting $\Delta m_{N_{1}} \equiv m_{N_{1}} - m_{\eta_{R}}$,  can be achieved, still leading to a correct thermal relic abundance  for $\Delta m_{N_{1}}$ below $\sim 20$ GeV. In addition, we go beyond ref.~\cite{Klasen:2013jpa} by identifying the existance of long-lived particles in this region. A small mass splitting between the new particles in the theory, namely, $\Delta m_{\eta N} \equiv  m_{\eta^{\pm}} - m_{N_{1}} $  and $\Delta m_{\eta^\pm} \equiv m_{\eta^{\pm}}  - m_{\eta_{R}}$, leads to regions in parameter space where either $N_{1}$ or $\eta^{\pm} $ can be long-lived particles , while also satisfying the dark matter relic abundance for $\Delta m_{\eta^\pm}$ below $\sim 1$ GeV. We study this phenomenological region that was often overlooked in the past in the minimal scotogenic model with thermal scalar dark matter, while respecting constraints from colliders, lepton number violation, electroweak precision observables, neutrino physics and direct detection searches. We also comment on differences in the scalar sector with the IHDM.


In addition, we discuss on the capabilities of future $e^{+} e^{-}$ colliders in detecting $N_{1}$, as detection of this new fermion may allow to distinguish the scotogenic from the IHDM scenario. As we will detail, the smallness of the yukawa couplings between $N_{1}$ and electrons leads to very small cross-sections, making the detection of a potential displaced vertex stemming from $N_{1}$ decaying to charged states very challenging. On the contrary, the detection of a long-lived $\eta^{\pm}$ could be possible in proton-proton collisions at the LHC with dedicated long-lived particle search strategies.

The rest of the paper is structured a follows. In section~\ref{sec:model} we review the particle content of the scotogenic model. In section~\ref{sec:darkmatter} we study the thermal scalar dark matter phenomenology, together with theoretical and experimental constraints imposed in our numerical scans. We then show in section~\ref{sec:SimResults} the allowed particle mass spectra consistent with a scalar dark matter candidate with the above constraints and the predictions of a long-lived $N_1$ and a long-lived $\eta^{\pm}$. We calculate cross-sections for selected benchmarks, and comment on future long-lived particle search strategies that could be performed at colliders, as well as on the reach with disappearing charged track searches for selected benchmarks. We then close in section~\ref{sec:summary}.

\section{The scotogenic model}
\label{sec:model}

We study a minimal extension of the SM with group $SU(3) \times SU(2)_{L} \times U(1)_{Y} \times \mathbb{Z}_{2}$, known as the scotogenic model~\cite{Ma:2006km}. This model has a new scalar doublet, $\eta$, and three singlet fermionic fields $N_{i}$, with $i = 1, 2, 3$. All these new particles are odd under $\mathbb{Z}_{2}$. The full particle content is shown at table~\ref{Table:particlecontent}.

\begin{figure}[h]
\centering
\includegraphics[width=0.35\textwidth,angle=0]{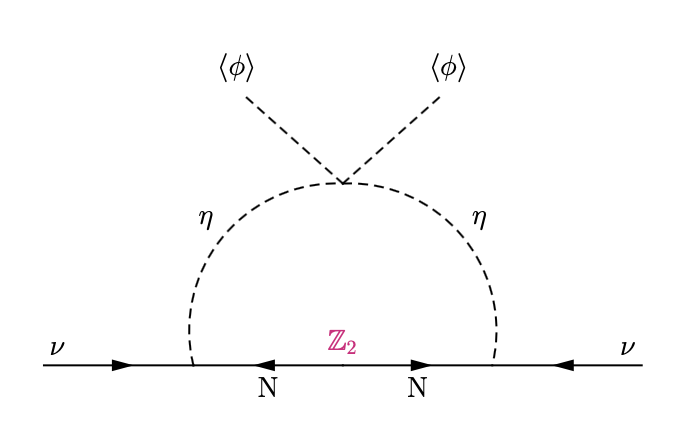}
\caption{One-loop diagram for the generation of neutrino masses in the scotogenic model.}
\label{fig:loop1}
\end{figure}

In this model, light neutrino masses are generated via the one-loop radiative seesaw mechanism, as shown in figure~\ref{fig:loop1}. The model also has two possible dark matter candidates: a bosonic neutral particle or a heavy fermion. We focus on the former scenario. 

We consider our DM candidate as a weakly interactive massive particle (WIMP) in the freeze-out scenario. We fix its mass as the lightest scalar neutral particle of the doublet, $\eta$, which has both a real ($\eta_{R}$) and imaginary components ($\eta_{I}$). For simplicity, we choose $\eta_R$ as the DM candidate. Since $\mathbb{Z}_{2}$ is an exact symmetry, $\eta_R$ does not acquire a vacuum expectation value.\\

\begin{table}[htb!]  
\centering
\begin{tabular}{| c| c | c | c | c | c | c |}
\hline
& \multicolumn{3}{ c |   }{\quad Standard Model \quad} &  \multicolumn{1}{ c |}{\quad Fermions \quad}  & \multicolumn{1}{ c | }{\quad Scalar \quad}  \\
\hline
           & \quad  $L$   & \quad $e$   & $\phi$   & \quad     $N$    & \quad $\eta$  \\
\hline                                                                            
$SU(2)_L$  & \quad   2    & \quad  1    &    2     & \quad      1    & \quad    2      \\
$Y$        & \quad  -1    & \quad -2    &    1     & \quad       0    & \quad    1        \\
$\mathbb{Z}_{2}$      & \quad  $+$   & \quad $+$   &   $+$    & \quad   $-$   & \quad   $-$    \\
$l$        & \quad   1    & \quad  1    &    0     & \quad    1        & \quad    0         \\
\hline
\end{tabular}
\caption{Particle content in the scotogenic model and their quantum numbers.}
\label{Table:particlecontent}
\end{table}

The scalar potential for the scotogenic model is

\begin{eqnarray}\label{Eq:ScalarPot}
V&=& m_{\phi}^{2}\phi^{\dag}\phi+m_{\eta}^{2}\eta^{\dag}\eta+\lambda_{1}(\phi^{\dag}\phi)^{2}+\lambda_{2}(\eta^{\dag}\eta)^{2}\nonumber\\
&+&\lambda_{3}(\phi^{\dag}\phi)(\eta^{\dag}\eta)
+\lambda_{4}(\phi^{\dag}\eta)(\eta^{\dag}\phi)\nonumber\\
&+&\frac{\lambda_5}{2}\left((\phi^{\dag}\eta)+(\eta^{\dag}\phi)\right)^{2},
\end{eqnarray}

where $m_{\phi}^{2}$, $m_{\eta}^2$ and $\lambda_{1-5}$ are real parameters~\cite{Ginzburg:2004vp, Ginzburg:2010wa}. In order to have a stable minimum of the potential at tree level, the following conditions are necessary~\cite{Ivanov:2006yq}
\begin{equation}
\lambda_{1},\lambda_{2}>0;\,\,\,\,\,\,\,\,\,\,\,\,\,\,\lambda_{3},\lambda_{3}+\lambda_{4}-|\lambda_{5}|> -2\sqrt{\lambda_{1}\lambda_{2}}.
\label{eq:lambdas_cond}
\end{equation}
Additionally, to comply with the requirement that the expansion of the scalar potential around its minimum is perturbatively valid, the scalar quartic couplings in equation~(\ref{Eq:ScalarPot}) are $\leq 1$. 

New interaction terms are present in the Lagrangian 

\begin{equation}
\mathcal{L}_{int}\subset -Y_{N}^{\alpha\beta}\bar{N_{\alpha}}\tilde{\eta}^{\dag}{L}_{\beta}-\frac{1}{2}\bar{N^{\alpha}}M_{\alpha\beta}N^{\beta \, c} +h.c.
\end{equation}
Here, $Y_{N}$ in the first term are the new Yukawa couplings, and is the only interaction between the new particles and the SM leptons. The last term corresponds to a Majorana mass with $M_{\alpha\beta}$ as a mass matrix.\\
After spontaneous symmetry breaking, the physical charged and neutral fields acquire a mass given by
\begin{eqnarray}
m_{\phi}^2 &=& 2\lambda_1 v^{2},\\
m_{\eta^{\pm}}^2 &=& m_{\eta}^2 + \frac{\lambda_3}{2}v^{2},\\
m_{\eta_{R}}^2 &=&m_{\eta}^2 + \left(\lambda_3+\lambda_4 + \lambda_5\right)\frac{v^2}{2},\\
m_{\eta_{I}}^2 &=& m_{\eta}^2 + \left(\lambda_3+\lambda_4 - \lambda_5 \right)\frac{v^2}{2}.
\label{Eq:etamasses}
\end{eqnarray}
To fix the lightest neutral particle of the inert doublet field, $\eta$, as the dark matter candidate at tree level, we have imposed
\begin{eqnarray}\label{Eq:ConditionsDM}
\lambda_4 +\lambda_5 &<& 0,\\
\lambda_5 &<&0,
\label{eq:lambda_5}
\end{eqnarray}
which set the DM candidate to be neutral, and choose $\eta_R$ (between $\eta_R$ and $\eta_I$) to be the dark matter candidate for definiteness.

\subsection{Neutrino mass generation}

To satisfy constraints from neutrino physics, we follow the original framework first proposed in~\cite{Ma:2006km}. After electroweak symmetry breaking, the resulting mass matrix of the light neutrinos at one-loop is 
\begin{eqnarray}
\mathcal{M}^{\nu}_{\alpha\beta}&=&\frac{Y^{N}_{\alpha i}Y^{N}_{\beta i}}{32\pi^{2}}m_{N_{i}}\left[\frac{m^{2}_{\eta_R}}{m^{2}_{\eta_R}-m_{N_{i}}^{2}}\ln\left({\frac{m^{2}_{\eta_R}}{m_{N_{i}}^{2}}}\right) \right. \nonumber\\
  &-& \left.\frac{m^{2}_{\eta_I}}{m^{2}_{\eta_I}-m_{N_{i}}^{2}}\ln\left({\frac{m^{2}_{\eta_I}}{m_{N_{i}}^{2}}}\right)\right].
\label{Eq:neutrinomassScoto}
\end{eqnarray}

where $m_{\eta_R}$ and $m_{\eta_I}$ are the masses of $\eta_R$ and $\eta_I$, respectively, and $N_i$ where $i=1,2,3$ are the masses of the new heavy fermions. Equation~(\ref{Eq:neutrinomassScoto}) has the following structure
\begin{equation}
\mathcal{M}^{\nu}_{\alpha\beta}=(Y^{N,T}\Lambda Y^{N})_{\alpha\beta}, 
\end{equation}

and $\Lambda$ given by

\begin{eqnarray}
\Lambda_{i}\!&=&\!\frac{m_{N_{i}}}{32\pi^{2}}\!\left[\frac{m^{2}_{\eta_R}}{m^{2}_{\eta_R}-m_{N_{i}}^{2}}\ln\left({\frac{m^{2}_{\eta_R}}{m_{N_{i}}^{2}}}\right) \right. \nonumber\\
  &-& \left.\frac{m^{2}_{\eta_I}}{m^{2}_{\eta_I}-m_{N_{i}}^{2}}\ln\left({\frac{m^{2}_{\eta_I}}{m_{N_{i}}^{2}}}\right)\right].
\end{eqnarray}

with

\begin{equation}
\Lambda =\left(\begin{array}{ccc}
\Lambda_1 & 0 & 0\\
0 & \Lambda_2 & 0\\
0 & 0 & \Lambda_3
\end{array}\right).
\end{equation}

Using the Casas-Ibarra parametrization~\cite{Casas:2001sr}, the matrix for the new Yukawa terms can be written as
\begin{equation}
Y^{N}=\sqrt{\Lambda^{-1}}R_{\text{rot}}\sqrt{m_{\nu}}U^{\dag}_{\text{PMNS}},
\end{equation}

where $R_{\text{rot}}$ is a random complex orthogonal matrix, $U_{\text{PMNS}}$ is the Pontecorvo-Maki-Nakagawa-Sakata matrix and $m_{\nu}$ is the diagonalized neutrino mass matrix. All light neutrinos acquire mass radiatively, see fig.~\ref{fig:loop1}.

\section{Scalar dark matter in the scotogenic model}
\label{sec:darkmatter}

We study the DM phenomenology of the scotogenic model considering the neutral scalar particle, $\eta_R$, as the DM candidate in the freeze-out scenario. We perform a detailed numerical scan for the model parameters with various experimental and theoretical constraints, to be detailed below. We explore a mass region for the dark matter candidate below $500$ GeV. This is motivated by the following. As opposed to the Inert Higgs Doublet model (IHDM)~\cite{Deshpande:1977rw}, the scotogenic model has extra contributing annihilation channels involving new fermions, allowing to decrease the relic density value. Still this is not enough to produce sizable differences between the IHDM and the scotogenic model for DM masses above 500 GeV~\cite{Borah:2017dfn} (see also fig.~\ref{fig:Relic_density}). However, we find from our numerical calculation using {\tt Micromegas 5.0.2}~\cite{Belanger:2014vza} that the presence of a new light fermion, with a mass close to the DM mass, affects the total relic density for dark matter masses below 500 GeV. In order to understand the nature of this result, we can study in a simplify way the equations that govern the dark matter freeze-out. Such equations govern changes in the relic density, as it is given by
\begin{equation}
\Omega_{\text{DM}}=\left[\frac{4\pi^{3}G g_{\ast}(m)}{45}\right]^{1/2}\frac{x_{F}T_{0}^{3}}{3-\langle\sigma v\rangle\rho_{\text{cr}}},
\end{equation}
where $x_{F}=m/T$ is a co-moving variable for the freeze-out time, $\rho_{\text{cr}}$ is the critical density of the Universe, $T_{0}$ the photon temperature today and $\langle\sigma v\rangle$ is the thermal average annihilation cross section~\cite{Dodelson:2003ft}. The number of degrees of freedom is defined as
\begin{equation}
g_{\ast } =\sum\limits_{bosons} g_{i} \frac{T_{i}^{4}}{T^{4}}+ \frac{7}{8}\sum\limits_{fermions}g_{i}\frac{T_{i}^{4}}{T^{4}} +g_{\ast ,NR}
\label{eq:grados_lib}
\end{equation}
where $g_{\ast ,NR}$ corresponds to the non relativistic term where the new particles of the model including the dark matter particle are contributing. In order to calculate those contribution, we have to compute the energy density due to $\rho\approx g_{\ast}T^{4}$~\cite{Kolb:1990vq}, where
\begin{equation}
\rho(T)=\frac{g}{(2\pi)^{3}}\int d^{3}p\frac{E}{e^{E/T}\pm 1}.
\end{equation}
For the non relativistic terms $T \ll m$~\cite{Bauer:2017qwy} and after some approximations we can find that $g_{\ast,\text{NR}}\propto \sum_{i} g_{i}e^{-m_{i}/T_{i}}\frac{m_{i}}{T_{i}}$ where $i$ represents each non-relativistic new particle in the scotogenic model. The mass of these new particles will contribute to this term, as is discussed in~\cite{Klasen:2013jpa}, given values that can affect the total relic density when the mass difference between the new light fermion and the other inert particles is small enough.
In what follows, we present details of our numerical analysis and show the parameter space found that allows to satisfy a correct relic abundance for DM masses $300 \lesssim m_{\eta_{R}} \lesssim 500$ GeV.

\subsection{Constraints}
\label{subsec:const}

{\emph{\textbf{Theoretical constrains:}}} The theoretical constraints considered include restriction on model parameters such that we have a stable DM candidate, and conditions respecting perturbativity on the quartic couplings.
As discussed in section~\ref{sec:model}, to ensure that the scalar potential is bounded from below at tree level, we apply conditions to the $\lambda_i$ (where $i=1-5$) parameters, given in equations~(\ref{eq:lambdas_cond}). We also impose the extra condition in eq.~(\ref{eq:lambda_5}) over $\lambda_5$, to fix $\eta_R$ as the DM candidate. We note this consideration does not ensure that the mass of $\eta_R$ will be the lightest particle if one considers mass corrections at one-loop level, which we do not consider. We explicitly require in our numerical scans to consider only points where $\eta_{R}$ is the lightest particle.

Another important theoretical consideration is to preserve the $\mathbb{Z}_2$ symmetry at low energy scales in order to keep the stability of the DM. Despite the fact that the $\mathbb{Z}_2$ symmetry is conserved at tree level, it can be broken at one loop, if we consider the running of the renormalization group equations (RGE). For our selected benchmarks (to be described below), we cross-check that the $Z_2$ symmetry is still unbroken under RGE evolution~\cite{Merle:2015ica}.

\textbf{\emph{Experimental constraints}}: The experimental constraints we consider in our analysis are: data from neutrino oscillation parameters, electroweak precision observables, restrictions from flavor violating processes, collider constraints, dark matter relic abundance and direct detection.

\begin{itemize}
\item \textbf{Neutrino oscillations:} The scotogenic model provides a mechanism to explain neutrino mass generation in the SM. For this purpose, new Yukawa couplings are introduced which are related to the mass differences of the SM neutrinos. We consider normal mass ordering, and let the mass of the lightest neutrino to vary between $10^{-13}$ eV and $9\times10^{-10}$ eV. We fix the mixing angles and squared mass differences according to~\cite{deSalas:2017kay}, so that our analysis is compatible with neutrino oscillation measurements.

\item \textbf{Electroweak precision tests}. We impose constrains on the $T$  parameter, as it is constrained by the splitting between the neutral and charged components of the new scalar fields. We require $-0.00022\leq \delta\rho\leq 0.00098$ at $3\sigma$ to have consistence with electroweak precision data~\cite{Zyla:2020zbs}.

\item \textbf{Lepton flavour violation (LFV):}  We apply rigorous limits on some rare processes that can occur in the scotogenic model~\cite{Toma:2013zsa}: BR$(\mu \to e \gamma) < 4.2 \times 10^{-13}$~\cite{TheMEG:2016wtm},  BR$(\mu \to e e e)  < 1. \times 10^{-12}$~\cite{Bellgardt:1987du}, CR$(\mu^-, {\rm Au} \to e^-, \rm Au) < 7 \times 10^{-13}$~\cite{Bertl:2006up}.

\item \textbf{Colliders constraints}: Following~\cite{Tanabashi:2018oca}, we consider $m_{\eta^\pm} \geq 100 \,{\rm GeV}$ and $122~{\rm GeV} \leq m_{h^0} \leq  128$ GeV. When the mass of $\eta_R$ is smaller than the mass of the SM Higgs, new channels for the Higgs boson going to invisibles opens up. Therefore, we also impose $BR(h^{0}\rightarrow \text{inv})\leq 19\%$~\cite{CMS:2018yfx}, $BR(h^{0}\rightarrow \gamma\gamma)/BR(h^{0}\rightarrow \gamma\gamma)_{\text{SM}}\gtrsim 0.84$ and $BR(h^{0}\rightarrow \gamma\gamma)/BR(h^{0}\rightarrow \gamma\gamma)_{\text{SM}}\lesssim 1.41$~\cite{Zyla:2020zbs}.

\item\textbf{Dark matter}: We assume a cold dark matter scenario considering a WIMP like DM particle. The relic density for a single dark matter candidate must fulfill the cosmological limits derived by Planck satellite data: $\Omega h^{2} = 0.1200\pm 0.0036 (3\sigma)$~\cite{Aghanim:2018eyx}. Also, our DM scenario can be proved by direct detection (DD) searches. In this work, we apply limits set by the XENON1T experiment on WIMP-nucleon spin independent elastic scattering cross section~\cite{Aprile:2018dbl}. These limits and the analysis details for the calculation of the DM relic abundance are explained in the next subsection.
\end{itemize}

\subsection{Numerical analysis}

For the implementation of the scotogenic model with the above constraints, we use {\tt{SARAH 9.3.1}}~\cite{Staub:2013tta,Staub:2015kfa}. We use {\tt SPHENO 4.0.3}~\cite{Porod:2003um,Porod:2011nf} and {\tt FlavourKit}~\cite{Porod:2014xia} to calculate decays and the physical particle spectrum with the files generated by {\tt SARAH} as input. The montecarlo analysis was made in python following the ranges of the parameters that are listed in table~\ref{Table:RunI}.
Dark matter abundance, and dark matter-nucleon spin-independent cross section at tree level are calculated with {\tt Micromegas 5.0.2}~\cite{Belanger:2014vza}. 

\begin{table}[htb!]
\centering
\begin{tabular}{| c | c  | }
\hline
{Parameter}& \multicolumn{0}{ c |   }{\quad Scanned range }   \\
\hline
$\lambda_{1}$ & \quad [$10^{-8}$ , $1$]         \\                                                                            
$\lambda_{2}$  & \quad  [$10^{-8}$ , $1$]                 \\
$\lambda_{3}$        &   $\pm$[$10^{-8}$ , $1$]                 \\
$\lambda_{4}$        &   $\pm$[$10^{-8}$ , $1$]          \\
$\lambda_{5}$        &   $\pm$[$10^{-8}$ , $1$]                 \\
$m_{\eta}$ [GeV]       & \quad   [$10$ , $1000$]  \\
$M_{N_{1}}$ [GeV]       & \quad  [$50$ , $5000$]                   \\
$M_{N_{2}}$ [GeV]        & \quad  [$5\times10^3$ , $2\times10^6$]                   \\
$M_{N_{3}}$ [GeV]        & \quad  [$5\times10^3$ , $3.5\times10^6$]                 \\
\hline
\end{tabular}
\caption{Input parameters used in our numerical scan.}
\label{Table:RunI}
\end{table}

\subsection{Relic Density and Direct searches}

\begin{figure}[h]
\centering
\includegraphics[width=0.5\textwidth,angle=0]{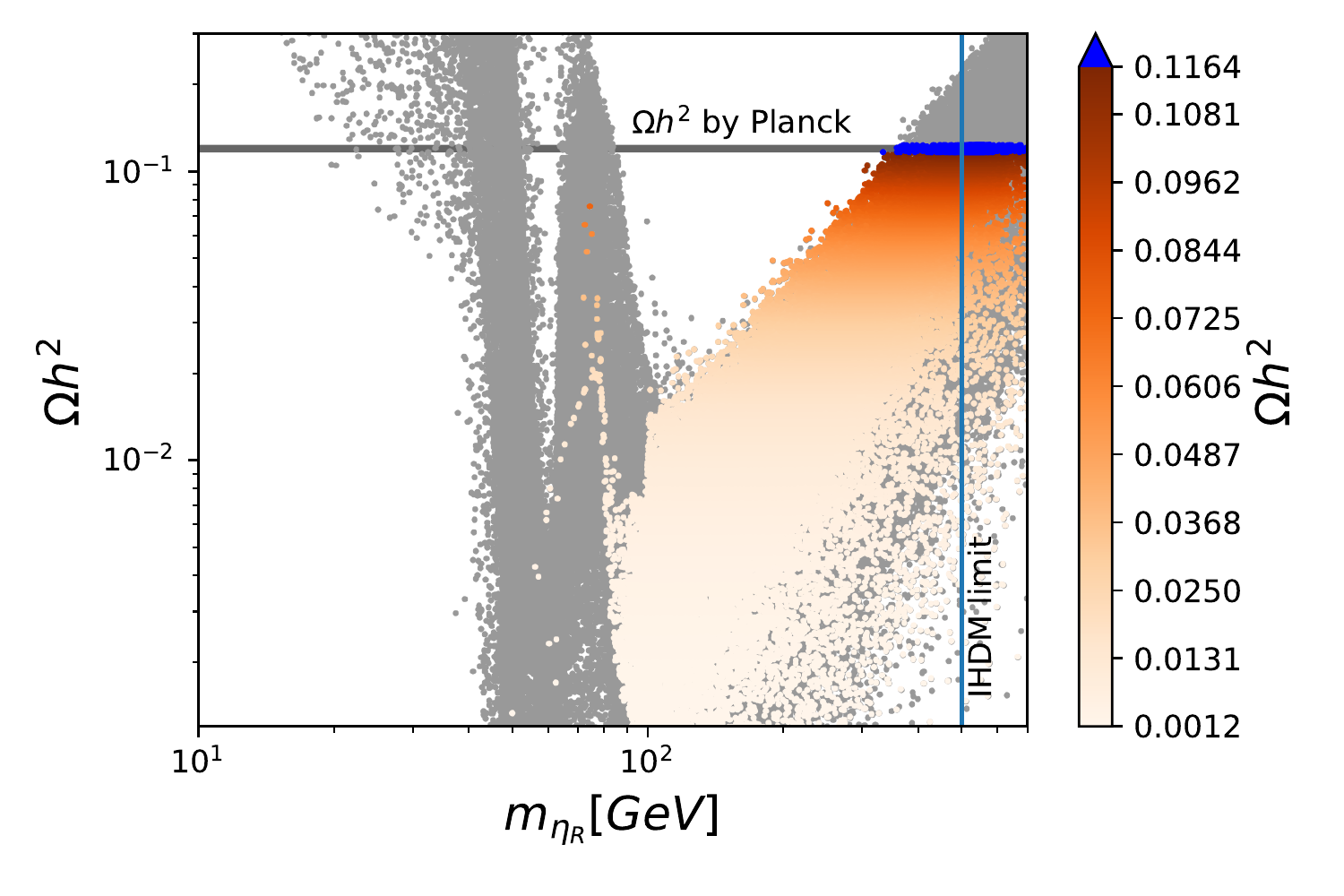}
\includegraphics[width=0.5\textwidth,angle=0]{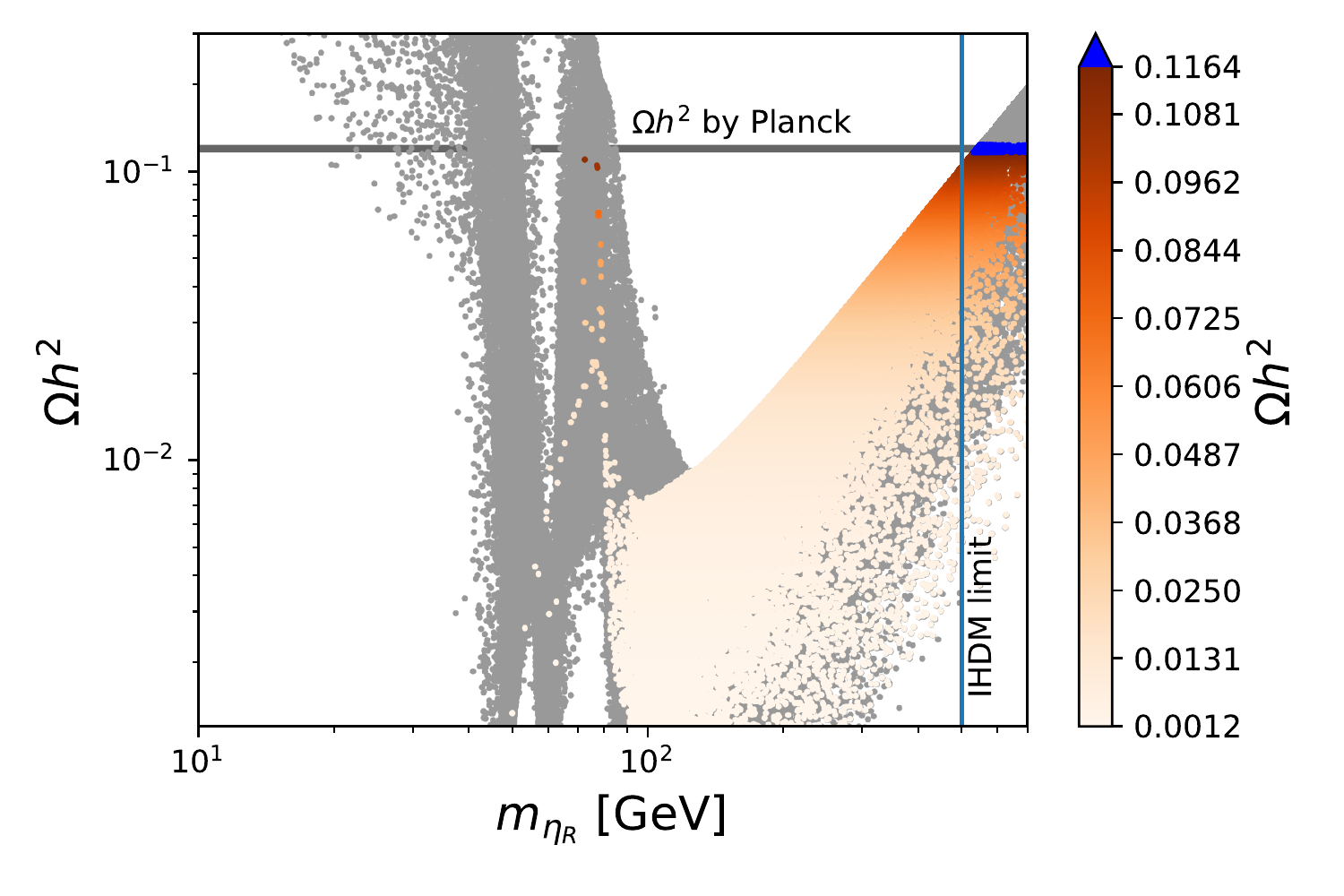}

\caption{ Relic density $\Omega_{\eta_{R}}h^2$ as a function of the DM mass, $m_{\eta_R}$, for the scotogenic model (top) and the IHDM (bottom). Grey points are results that are excluded when applying the constraints described in the text. Blue points fall within the $3\sigma$ C.L. band measured by Planck~\cite{Aghanim:2018eyx} for a cold dark matter scenario. The vertical line at $m_{\eta_{R}} = 500$ GeV represents the IHDM limit, emphasizing this model can not explain a relic abundance below the line. Points in the orange color bar are solutions with different fractions of the relic density, and are allowed by the experimental direct searches and constraints described in the main text.}
\label{fig:Relic_density}
\end{figure}

For the DM relic abundance calculation, we consider annihilation and co-annihilation of $\eta_R$ into bosons and fermions. Figure~(\ref{fig:Relic_density}) shows the dark matter relic abundance as a function of the DM mass, $m_{\eta_R}$, for both the IHDM and the scotogenic model. The dark gray horizontal band corresponds to the relic density for cold dark matter as measured by Planck at $3\sigma$~\cite{Aghanim:2018eyx}. The blue dots are results falling on the Planck band in both models, which can explain the total relic abundance present in the Universe. For the scotogenic model, these points are located at DM masses approximately between $300-900$ GeV. The orange dots are points where $\eta_R$ is a sub-dominant candidate for DM and therefore, another contribution to DM would be required. Their intensity decreases according to the fraction of the relic density they are able to explain. Grey points are results that are excluded when applying the constraints described in the text. Orange and blue points in the figures satisfy the theoretical and experimental constraints mentioned in section~\ref{subsec:const}.

We note that in fig.~(\ref{fig:Relic_density}) there is a first drop at a mass $~60$ GeV, which corresponds to the $Z$ boson peak when $m_{\eta_{R}}\sim M_{Z}/2$. The second drop comes from the Higgs contribution, when $m_{\eta_{R}}\sim m_{h^0}/2$. Both annihilation channels are via the s-channel. For masses around $~90$ GeV, quartic interactions become important, and $\eta_R$ can annihilate into $W^{+}W^{-}$, which is why we see a third drop. In addition, co-annihilation of $\eta_R$ with $\eta_I$ and $\eta^{\pm}$ may appear in regions of the parameter space where the mass splittings are small. Depending on the mass of $N_1$, we can find annihilation of the dark matter into neutrinos through $N_1$ in the t-channel. 

\begin{figure}[h]
\centering
\includegraphics[width=0.5\textwidth,angle=0]{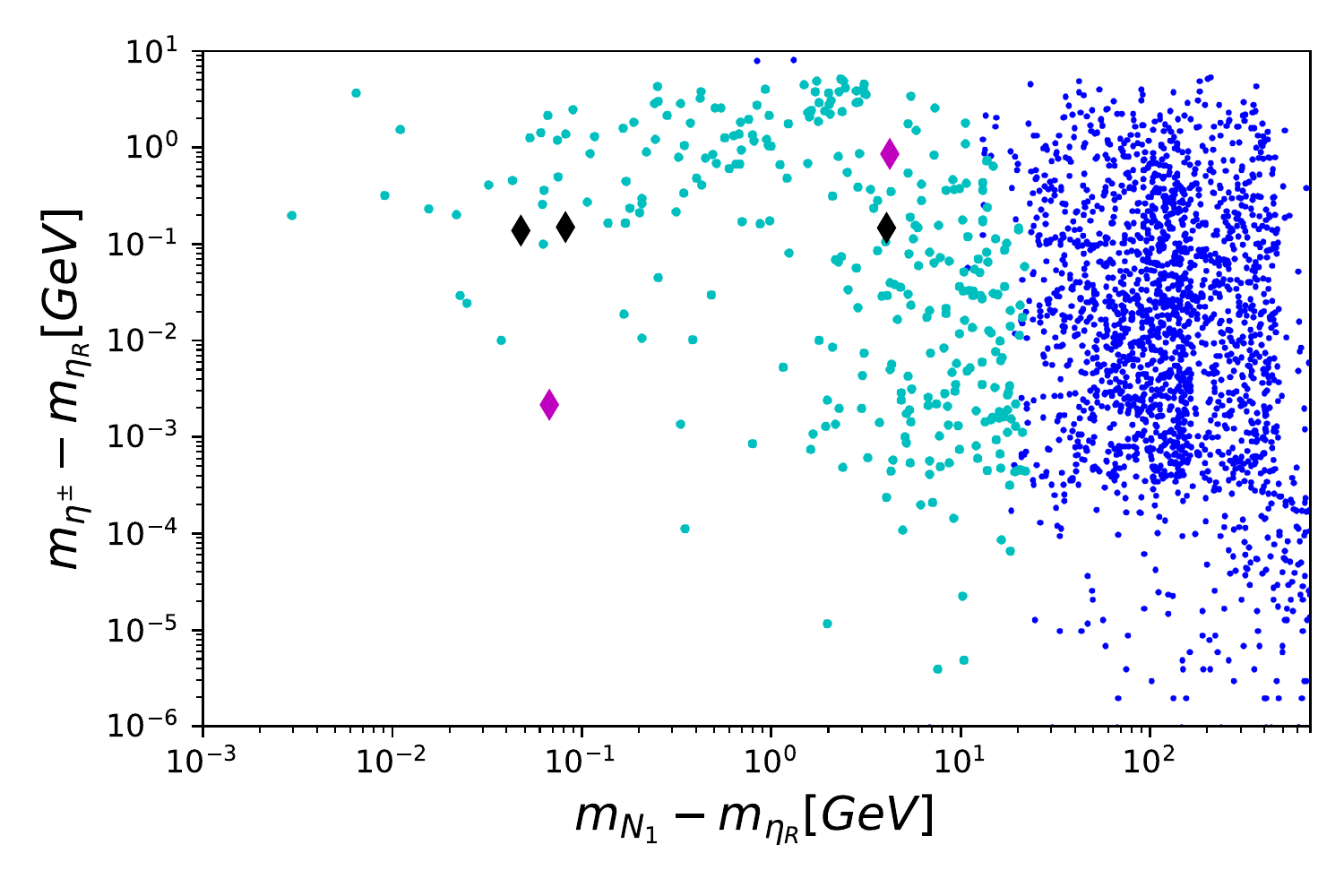}
\caption{ Mass difference $\Delta m_{N_{1}} \equiv m_{N_{1}} - m_{\eta_{R}}$ as a function of $\Delta m_{\eta^\pm} \equiv m_{\eta^{\pm}}  - m_{\eta_{R}}$ for points satisfying the total relic abundance in the scotogenic model. Blue points correspond to mass differences when fixing $m_{\eta_{R}}> 500$ GeV, while cyan ones correspond to  $m_{\eta_{R}}< 500$ GeV. Diamonds correspond to the benchmarks B1 and B2 in magenta (see table~\ref{Table:N1BMs}) and B3, B4 and B5 in black (see table~\ref{Table:etaBMs}).}
\label{fig:Relic_massDiff}
\end{figure}

As was mentioned in section~\ref{sec:darkmatter}, we can find a region below the DM mass of $500$ GeV with accepted results that contributes to the total relic density. These allowed regions are harder to find in our scans, and are the consequence of the small mass difference between the new particles of the inert sector, as shown in fig.~(\ref{fig:Relic_massDiff}). We can see that for DM masses below 500 GeV (cyan points) and for $\Delta m_{N_{1}}$ below $\sim 20$ GeV, the totality of the relic abundance can be satisfied 
while $\Delta m_{\eta^\pm} \lesssim 1$ GeV.

Our dark matter scenario can be proved with direct searches for DM. We calculate the the spin-independent cross-section from dark matter scattering off nucleons at tree level, which can be constrained by deep underground experiments like XENON1T~\cite{Aprile:2018dbl}. The $\eta_R$-nucleon spin independent elastic scattering cross-section is mediated through a Higgs boson $h^0$, and a $Z$ boson. The interaction through the Higgs channel is the dominant contribution across the entire parameter space. This is because interactions with $Z$ bosons will depend on the mass splitting between $\eta_R$ and $\eta_I$.

\begin{figure}[h]
\centering
\includegraphics[width=0.5\textwidth,angle=0]{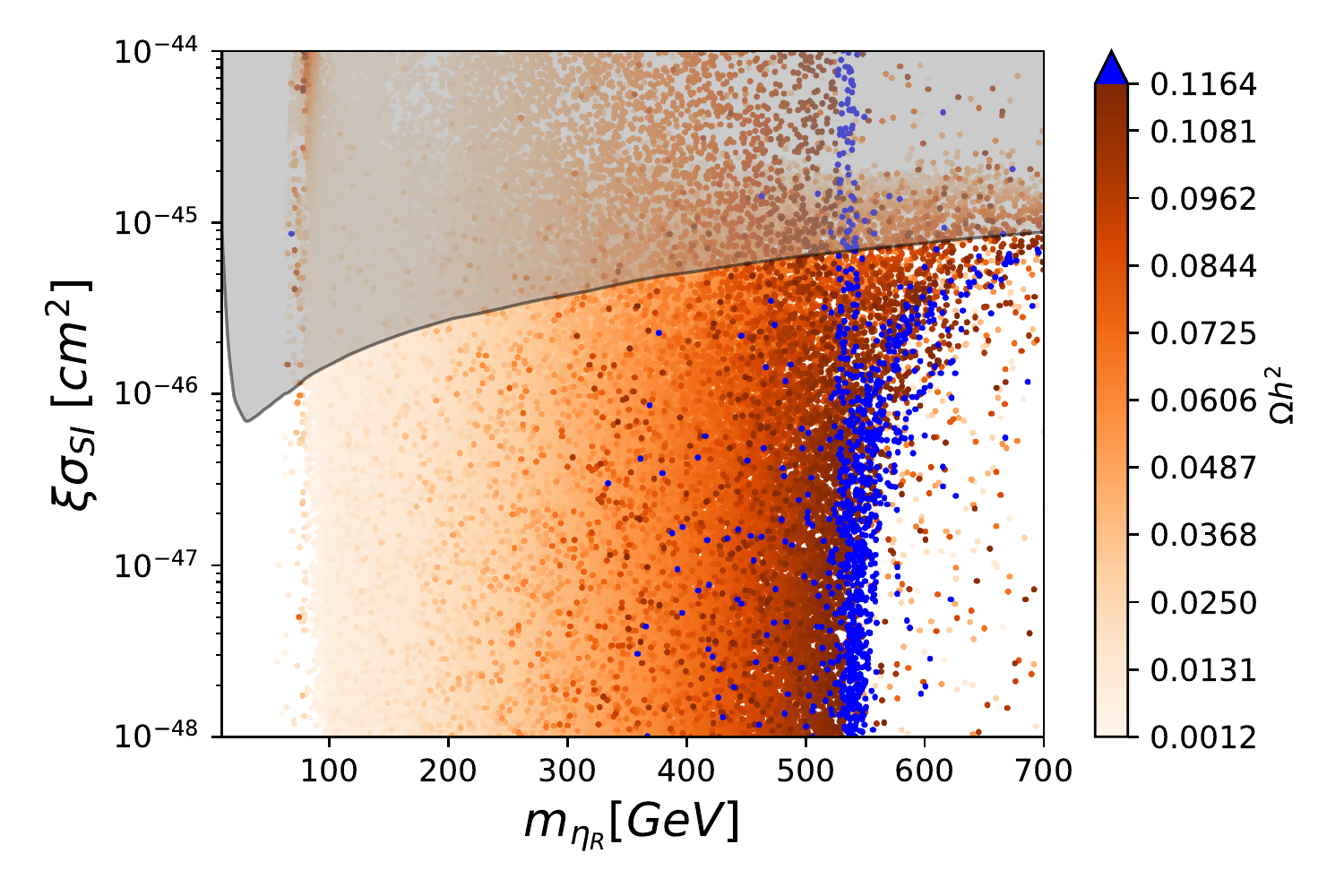}
\caption{$\eta_R$-nucleon spin independent elastic scattering cross-section as a function of $m_{\eta_R}$. The dark grey line denotes the upper bound from XENON1T~\cite{Aprile:2018dbl}. }
\label{fig:DD_etaR}
\end{figure}

Fig.~(\ref{fig:DD_etaR}) shows the constraints we obtain from direct searches in the scotogenic model. The vertical axis represent the $\eta_R$-nucleon spin independent elastic scattering cross-section. Each point in the plot is weighted by $\xi=\Omega_{\eta_R}/{\Omega_{\text{Planck}}}$.  In the horizontal axis, we have the mass of $\eta_R$.  The color bar represents the relic density value, where blue points can explain the total DM abundance. The dark gray line corresponds to the 95\%CL upper limit set by XENON1T on WIMP-nucleon spin independent elastic scattering cross-section~\cite{Aprile:2018dbl}. Points overlapping with the dark gray region are excluded in our model. Non-excluded points satisfying the totality of the relic abundance are shown in blue, with masses between $\sim 300-700$ GeV.

\begin{figure}[h]
\centering
\includegraphics[width=0.5\textwidth,angle=0]{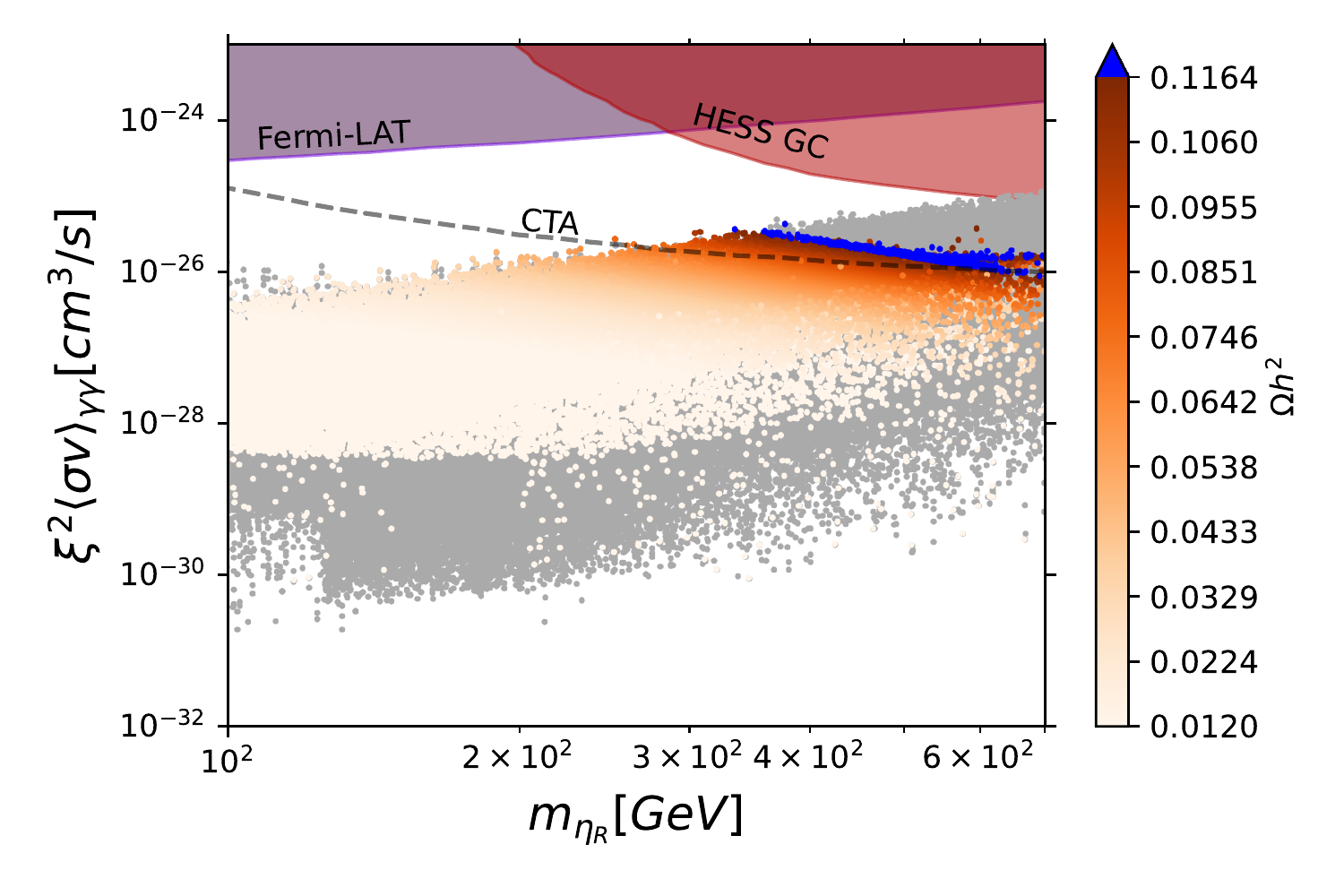}
\caption{Dark matter annihilation cross section – weighted
by the relative abundance – into $\gamma$ rays, for annihilation to $W^{+}W^{-}$. Dark purple and dark red regions represent the upper limit by Fermi-LAT~\cite{Fermi-LAT:2016afa} and H.E.S.S~\cite{HESS:2016mib} at $95\%$ C.L, respectively. The black dashed curve shows a sensitivity projection for CTA~\cite{CTAConsortium:2017dvg}.}
\label{fig:indirect_etaR}
\end{figure}

Indirect detection constraints in the scotogenic model for scalar dark matter masses below 500 GeV where previously considered in ref.~\cite{Klasen:2013jpa}. Neutrino signals from the annihilation of dark matter in the scotogenic scenario with scalar dark matter were studied recently in~\cite{deBoer:2021pon}. Here we study current constraints from indirect detection search experiments, focusing on the mass region between 100-700 GeV (where our benchmarks in table~\ref{Table:etaBMs} lie). Motivated by previous results obtained by reference~\cite{Avila:2019hhv}, we analyze the dark matter annihilation into $W^{+}W^{-}$, considering $\gamma$ rays as final states. We contrast our results with current limits imposed by the H.E.S.S telescope~\cite{HESS:2016mib} and the Fermi-LAT satellite~\cite{Fermi-LAT:2016afa}. Figure~\ref{fig:indirect_etaR} shows the dark matter annihilation cross sections versus $\eta_R$ mass. The $\eta_R$ annihilation cross sections is weighed by the correspondent branching ratio $\xi^{2}$. Blue points represent results satisfying a total relic density, while grey points are excluded by constraints detailed in the previous subsection. The dark red line represents the upper limit from H.E.S.S at $95\%$ C.L. considering an Einasto profile for dark matter when annihilating into $W^{+}W^{-}$, looking the galactic center (GC). In addition, the dark purple line shows the upper limit at $95\%$ C.L set by Fermi-LAT when observing dwarf spheroidal satellite galaxies (dSphs) in the Milky Way assuming annihilation into $W^{+}W^{-}$. We complement our analysis with sensitivity projections for future experiments such as the Cherenkov Telescope Array (CTA) when dark matter annihilate into $W^{+}W^{-}$ channel assuming an Einasto dark matter density profile looking the galactic halo~\cite{CTAConsortium:2017dvg}. Our points are beyond the reach of current experiments. Nevertheless, blue points satisfying a total dark matter abundance (including our benchmarks), could be tested by future experiments as CTA.

\section{Long-lived particles}
\label{sec:SimResults}

As detailed above, the small mass differences $\Delta m_{N_{1}}$ and $\Delta m_{\eta^\pm}$ allows a dark matter candidate with a mass $m_{\eta_{R}}$ below 500 GeV. This also motivates the study of long-lived particles (LLP) in the GeV mass range, as small mass differences between $\mathbb{Z}_{2}$-odd fermions and scalars, together with small Yukawa couplings, can lead to particles decaying with macroscopic lifetimes in this scotogenic scenario with a consistent scalar dark matter candidate (described in the previous section). We focus on two different scenarios according to the spectrum of the $\mathbb{Z}_{2}$-odd particles, leading to two possible long-lived particle candidates:
i) $m_{N_{1}} > m_{\eta^{\pm}} > m_{\eta_{R}} $, with $N_{1}$ as LLP and ii) $m_{\eta^{\pm}}$ as LLP, with $m_{\eta^{\pm}} > m_{N_{1}} > m_{\eta_{R}} $.  

\subsection{Long-lived $N_{1}$}

We first focus on a scenario where $N_{1}$ can be a long-lived particle, and decouple the other fermions $N_2$ and $N_3$ to be heavier, well above the TeV scale (see table~\ref{Table:RunI}). The possibility of long-lived fermions $N_2, N_3$ in the scotogenic scenario was address in ref.\cite{Hessler:2016kwm}, where the authors discuss LHC signatures of the heavier fermions decaying to $N_{1}$, which was a fermionic FIMP DM candidate. Since in this work we focus on a scalar DM candidate, if the mass of $N_{1}$, $m_{N_{1}}$, is heavier than the $\eta$ particles, then pair production of a long-lived $N_{1}$ by $e^{+}e^{-}$ annihilation trough $\eta^{\pm}$ can lead to an additional LLP signatures to evaluate.

In this scenario, $N_{k}$ (with $k=1$ in our case) can decay to $N_{k}\rightarrow l^{\mp}_{j}\eta^{\pm}$ and $N_{1}\rightarrow \nu_{j}  \eta_{\alpha}$ (with $j=1,2,3$ and $\alpha= R, I$), with partial decay widths given by~\cite{Ho:2013spa}

\begin{eqnarray}
\Gamma_{N_{k}\rightarrow l^{+}_{j}\eta^{-}}&=&\Gamma_{N_{k}\rightarrow l^{-}_{j}\eta^{+}}\nonumber\\
&=&\frac{|Y_{jk}|^{2}(m_{N_{k}}^{2}+m_{l_{j}}^{2}-m_{\eta^{+}}^{2})}{32\pi m_{N_{k}}^{3}}\nonumber\\
&\times & \!\!\sqrt{(m_{N_{k}}^{2}-m^{2}_{l_{j}}-m^{2}_{\eta^{+}})^{2}\!-\! 4m^{2}_{l_{j}}m^{2}_{\eta^{+}}}\\ 
\Gamma_{N_{k}\rightarrow \nu_{j} \eta_{\alpha}}&=& \Sigma_{j} \frac{|Y_{jk}|^{2}}{32\pi m_{N_{k}}^{3}} (m_{N_{k}}^{2}-m_{\eta_{\alpha}}^{2})^{2}.
\label{eq:N1decays}
\end{eqnarray}

\begin{figure}[h]
\centering
\includegraphics[width=0.45\textwidth,angle=0]{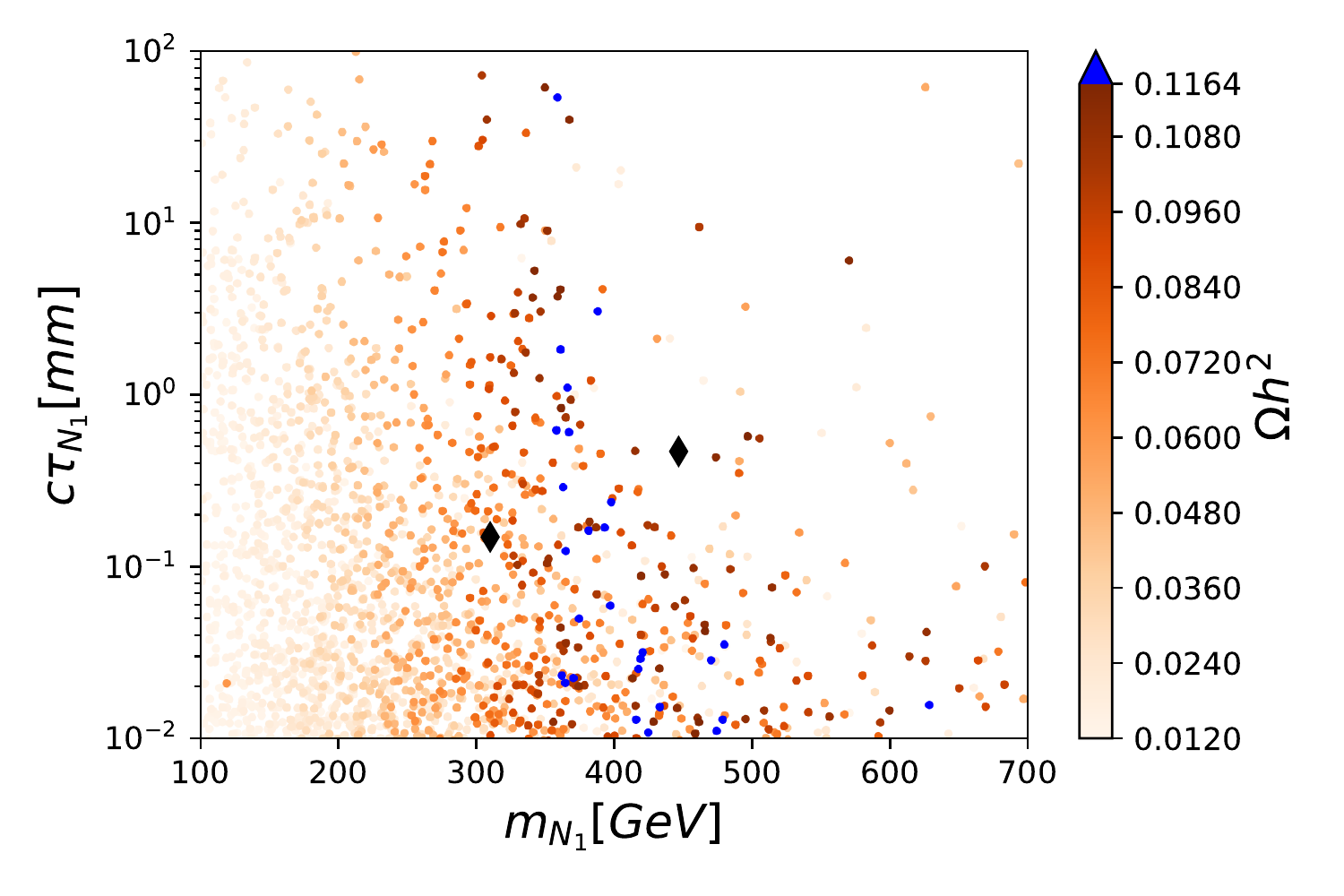}
\caption{Proper decay distance of $N_{1}$ as a function of mass, for different values of the relic density $\Omega_{\eta_{R}}h^2$.}
\label{fig:decayDistance}
\end{figure}

\begin{figure}[h]
\centering
\includegraphics[width=0.45\textwidth,angle=0]{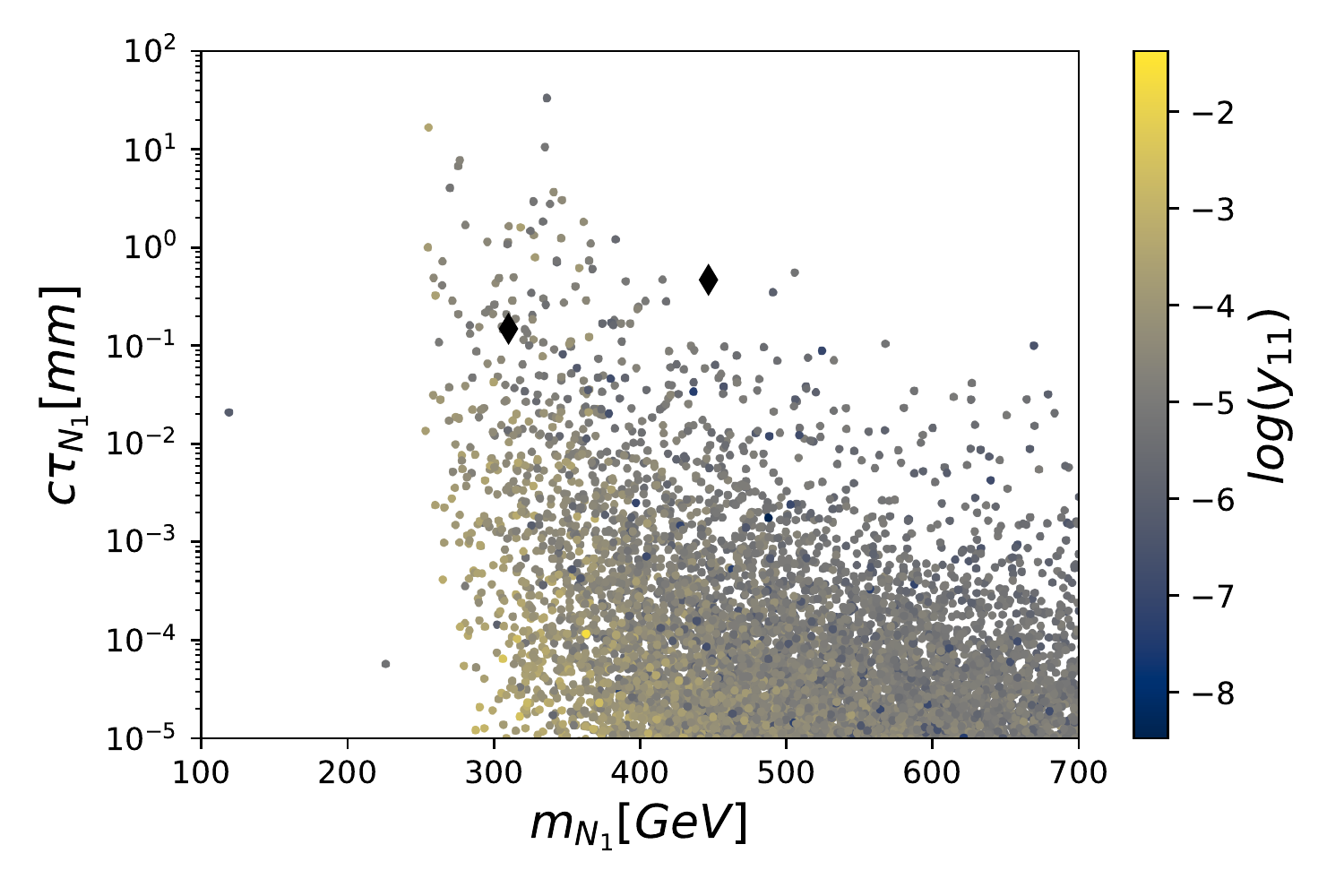}
\caption{Proper decay distance of $N_{1}$ as a function of mass, for different values of the yukawa coupling $y_{11}$. For B1 we have $y_{11}=4.91\times 10^{-6}$ and $y_{11}=5.45\times 10^{-4}$ for B2. All points satisfy at least the 50\% of the relic abundance. }
\label{fig:decayDistanceYukawa}
\end{figure}

Fig.~(\ref{fig:decayDistance}) shows the proper decay distance of the lightest fermion, $c\tau_{N_{1}}$, as a function of its mass. The blue dots are the ones that survive to all constraints described in section~\ref{subsec:const} and can also explain the total dark matter abundance. Orange dots represents the points that are not ruled out by any constraint, but can only satisfy a fraction of the DM relic density. We find benchmark points with proper decay distances of $\approx \mathcal{O}(1) $ mm in table~(\ref{Table:N1BMs}). These two benchmarks, B1 and B2 are the ones with total (B1) and partial (B2) relic abundance while also satisfying a macroscopic decay length of $N_{1}$ of a few millimeters (after considering the boost factor) and decaying to visible charged particles. As it can be seen from the table, production cross-sections for $N_{1}$ at $3$ TeV are low, making the detection of a long-lived $N_{1}$ at future $e^{+}e^{-}$ very unlikely. This can be understood due to the smallness of the yukawa coupling $y_{11}$ between $N_{1}$ and electrons, as shown in figure~\ref{fig:decayDistanceYukawa}. Sensitivities will be further affected by displaced vertex efficiencies and branching ratios of $N_{1}$ decaying to leptons and a charged $\eta^{\pm}$ (see eq.~(\ref{eq:N1decays})), which reaches $\sim 40 \%$ for B1 and $\sim 10\%$ for B2.

\begin{table}[htb!]  
\begin{tabular}{| c| c | c | c | c | }
\hline
 \multicolumn{1}{| c |}{\quad 	Parameter \quad} &   \multicolumn{1}{ c |}{\quad B1 \quad}  & \multicolumn{1}{ c | }{\quad B2 \quad}   \\
\hline                                                                    
$\lambda_3$  & \quad      $-2.809\times 10^{-4}$     & \quad   $2.322\times 10^{-8}$     \\
\hline
$\lambda_4$      & \quad      $ 1.16\times 10^{-5}$     & \quad   $-1.538\times 10^{-5}$    \\
\hline
$\lambda_5$     & \quad   $-2.511\times 10^{-2}$   & \quad  $-2.878\times 10^{-5}$   \\
\hline
$m_{\eta}^{2}$ [GeV]      & \quad   $1.966\times 10^{5}$   & \quad  $9.608\times 10^{4}$     \\
\hline
$m_{\eta_R}$ [GeV]      & \quad   $442.535$   & \quad  $309.961$    \\
\hline
$m_{\eta_I}$ [GeV]      & \quad   $444.252$   & \quad  $309.964$    \\
\hline
$m_{\eta^\pm}$ [GeV]      & \quad   $443.394$   & \quad  $309.964$  \\
\hline
$m_{N_1}$ [GeV]    & \quad   $446.754$    & \quad  $310.028$     \\
\hline
$c\tau_{N_1}$ [mm]    & \quad   $0.467$    & \quad  $0.149$    \\
\hline
$\sigma(e^{+}e^{-}\rightarrow N_{1}N_{1})$ [fb]    & \quad    $9.89\times 10^{-20}$   & \quad    $1.68\times 10^{-11}$    \\
\hline
$\Omega h^{2}$     & \quad   $0.122$   & \quad  $0.092$    \\
\hline
\end{tabular}
\caption{Relevant model spectrum and parameters for two representative benchmarks, B1 and B2, with $N_{1}$ as a long-lived particle. Cross-sections at $e^{+}e^{-}$ were calculated at $\sqrt{s}= 3$ TeV with \texttt{MadGraph}~\cite{Alwall:2014hca}.}
\label{Table:N1BMs}
\end{table}

Unfortunately, the detection of a long-lived $N_{1}$ at future $e^{+}e^{-}$ seems very unlikely due to a small production cross section. However $N_{1}$ could still be probed with prompt searches for charged leptons $+$ missing transverse momenta or in association with photons~\cite{Ho:2013spa,Ahriche:2018ger}. We now discuss on the scenario where $\eta^{\pm}$ can be a LLP and comment on future LHC searches.

\subsection{Long-lived $\eta^{\pm}$}

We find scenarios where $\eta^{\pm}$ has a macroscopic lifetime while satisfying the dark matter relic abundance in the scotogenic model with scalar dark matter. There are two cases to consider. If $ m_{\eta^{\pm}} > m_{N_{1}}$, then $\eta^{\pm}$ can decay to $N_{1}$ and a charged lepton, with a rate given by~\cite{Borah:2018smz} 

\begin{eqnarray}
\Gamma_{\eta^{\pm}\rightarrow N_{1}l^{\pm}}&=&\frac{Y_{l 1}^{2}(m^{2}_{\eta^{\pm}}-(m_{N_1}+m_{l})^{2})}{8 m_{\eta^{\pm}}\pi}\\
&\times &\!\! \sqrt{1-\left(\frac{m_{N_1}-m_{l}}{m_{\eta^{\pm}}}\right)^{2}}\sqrt{1-\left(\frac{m_{N_1}+m_{l}}{m_{\eta^{\pm}}}\right)^{2}} \nonumber
\label{eq:decayetalepton}
\end{eqnarray}

Observable decays of $\eta^{\pm}$ to $\eta_{R}$ and a pion are also possible if kinematically allowed with rate~\cite{Borah:2018smz} 

\begin{eqnarray}
\Gamma_{\eta^{\pm}\rightarrow \eta_{R}\pi^{\pm}}&=&\frac{f^{2}_{\pi}g^{4}}{m^{4}_{W}}\frac{(m^{2}_{\eta^{\pm}}-m^{2}_{\eta_{R}})^{2}}{512 {m_{\eta^{\pm}}}{m_{\pi}}}\\
&\times &\!\! \sqrt{1-\left(\frac{m_{\eta_{R}}-m_{\pi}}{m_{\eta^{\pm}}}\right)^{2}}\sqrt{1-\left(\frac{m_{\eta_{R}}+m_{\pi}}{m_{\eta^{\pm}}}\right)^{2}} \nonumber
\label{eq:decayetapion}
\end{eqnarray}
where $f_{\pi}$ is the form factor and $m_W$ the mass of the $W$ boson. 

\begin{figure}[h]
\centering
\includegraphics[width=0.45\textwidth]{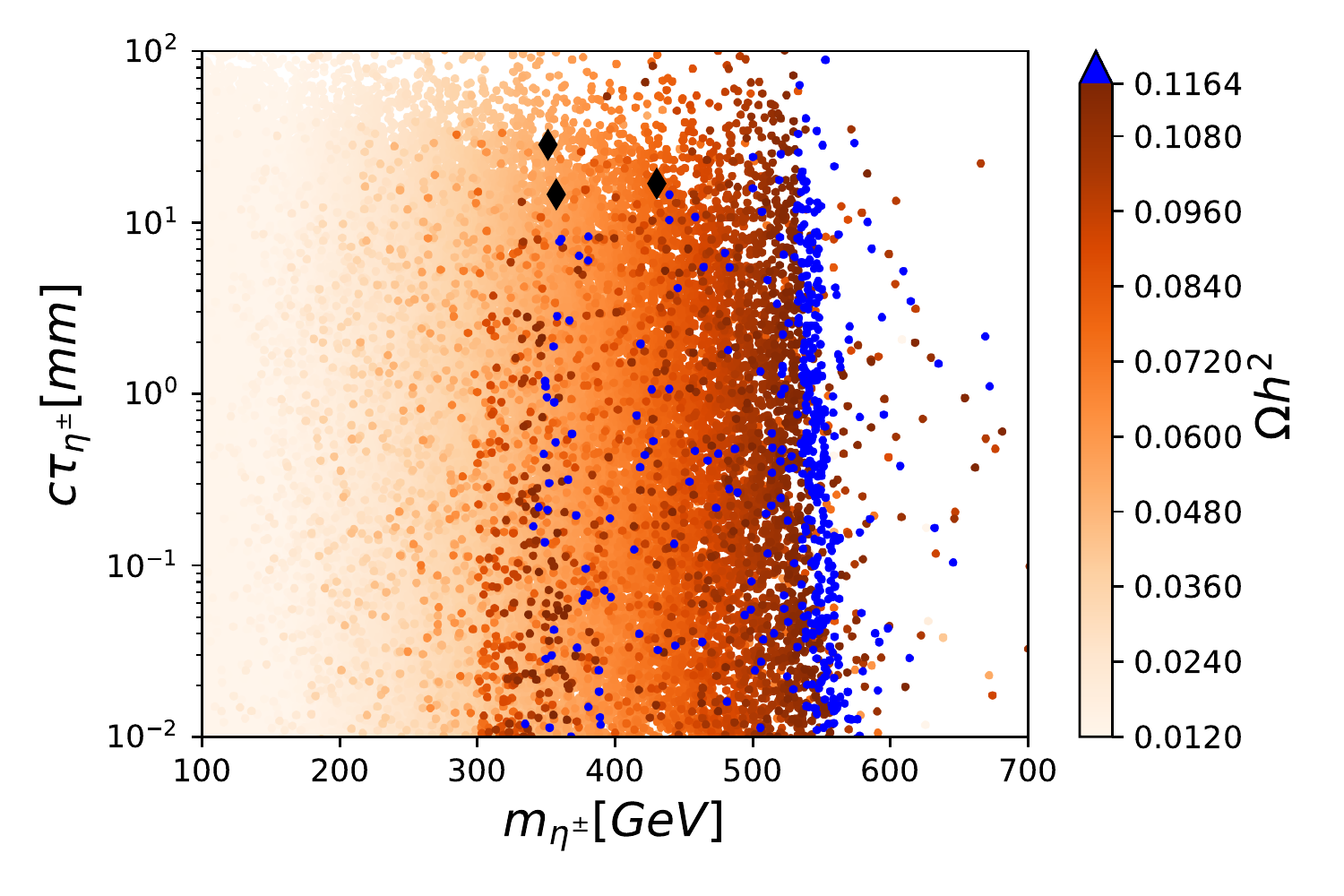}
\caption{Proper decay distance of  $\eta^{\pm}$ as a function of mass, for different values of the relic density $\Omega_{\eta_{R}}h^2$.}
\label{fig:decayDistanceChargedEta}
\end{figure}

Figure~\ref{fig:decayDistanceChargedEta} shows the proper decay distance of $\eta^{\pm}$ as a function of mass, for different values of the relic abundance satisfying all constraints from the previous section. Points in blue satisfy the total dark matter relic abundance, and are dominated by the decays of $\eta^{\pm}$ to DM and a pion, as the mass splitting $\Delta m_{\eta^\pm} \sim \mathcal{O}(100)$ MeV or below (see also figure~\ref{fig:Relic_massDiff}). 

We find representative benchmarks in table~\ref{Table:etaBMs} for proper decay distances of $\eta^{\pm}$ within the mm and cm range, for dark matter masses below $500$ GeV. In all benchmarks we find that the totality of the relic abundance can be satisfied, with  100$\%$ branching ratio of $\eta^{\pm} \rightarrow \eta_{R}\pi^{\pm}$ in B3 and B4, reaching$\sim 70\%$ in B5. In both cases, the search of a disappearing charged track signature at the LHC~\cite{ATLAS:2017oal,CMS:2020atg} is promising. Prospects for disappearing charged track searches were explored in the context of dark matter models with similar phenomenology in refs.~\cite{Belyaev:2016lok,Belyaev:2020wok} and in extensions of the scotogenic model recently in~\cite{Barman:2021ifu}.

We recast the ATLAS disappearing charged track search in ref.~\cite{ATLAS:2017oal}. The validation of this search was done in ref.~\cite{Chiang:2020rcv} by some of us, and here we use the same strategy. Events are selected with a missing transverse momenta trigger ($E^{miss}_{T}$), therefore, the signal topology contains a high$-p_{T}$ jet. Events are generated with \texttt{MadGraph}~\cite{Alwall:2014hca} at 14 TeV via a $pp \rightarrow \eta^{\pm}\eta^{\mp} j $ and $pp \rightarrow \eta^{\pm}\eta_{R} j $. We interface our events in \texttt{Pythia8}~\cite{Sjostrand:2014zea} and match them with up to two extra partons. A custom made detector simulation is performed inside \texttt{Pythia8} as in~\cite{Chiang:2020rcv}. The following analysis selections are imposed: $E^{miss}_{T} > 140$; no electrons or muons; at least one jet with $p_{T} > 140$\,GeV, and $\Delta\phi$ between the $E^{miss}_{T}$~vector and each of the up to four hardest jets with $p_{T}>50$\,GeV to be bigger than 1.0.

After these selections at the event level, we proceed with the selection of at least one short track or ``tracklet'', corresponding to the generator-level $\eta^{\pm}$ in our signal. This must satisfy: $p_{T}>20 $\,GeV and $0.1<|\eta|<1.9$; $122.5$ mm $<$ decay position $<295$ mm; $\Delta R$ distance between the tracklet and each of the up to four highest$-p_{T}$ jets with $p_{T}>50$\,GeV to be bigger than 0.4;  acceptance $\times$ efficiency map based on the above decay position and $\eta$ (provided by ATLAS in the auxiliary material 
of~\cite{ATLAS:2017oal}); tracklet with $p_{T}>$ 100\,GeV, correcting with an overall experimental efficiency factor of 0.57 as detailed in the ATLAS analysis\footnote{We assume these experimental efficiencies will remain the same at 14 TeV for the purpose of our estimations.}.

The disappearing track analysis efficiencies for our benchmarks ranges from  $0.2\%$ to about $0.6\%$, raising from B3 to B5. This is due to a larger decay length in B5. With  $3000/$fb of luminosity,  the total number of expected events for benchmarks that satisfy a total relic abundance can be seen in table~\ref{Table:etaBMs}, demonstrating that the high-luminosity LHC could test these scenarios. 

We finish this discussion by commenting that decays of $\eta^{\pm} \rightarrow N_{1} l$, with $l=e,\mu$ con also happen with significant branching fraction. But generally we find these points do not satisfy the totality of the relic abundance as $m_{\eta_{R}}\approx 200$ GeV (and $\Delta m_{\eta N}\approx 140$ MeV), which is very close to the smallest values of DM masses able to satisfy the totality (see figure~\ref{fig:Relic_density}). Nevertheless, such channels  could still provide the possibility of a novel collider signature when $\eta^{\pm}$ is pair produced and decays to a pair of displaced (soft) leptons in association with missing transverse momenta that could potentially be detected inside the inner trackers of the LHC experiments~\cite{Blekman:2020hwr}, complementing existing proposals for the detection of $\eta^{\pm}$ in the scotogenic model.  For mass differences  of $\Delta m_{\eta N} \sim \mathcal{O}(100)$ GeV, prospects for detecting prompt di-leptons + missing energy from the decay of $\eta^{\pm}$ produced at the LHC were studied previously in ref.~\cite{Baumholzer:2019twf}.

\begin{table}[htbp]  
\resizebox{\columnwidth}{!}{\begin{tabular}{| c| c | c | c | c | c |}
\hline
 \multicolumn{1}{| c |}{\quad 	Parameter \quad} &   \multicolumn{1}{ c |}{\quad B3 \quad}  & \multicolumn{1}{ c | }{\quad B4 \quad} & \multicolumn{1}{ c | }{\quad B5 \quad} \\
\hline                                                                    
$\lambda_3$  & \quad      $-2.392\times 10^{-5}$     & \quad  $ 3.305 \times 10^{-6} $  & \quad   $ 4.447 \times 10^{-5}$     \\
\hline
$\lambda_4$      & \quad      $-6.923\times 10^{-7}$  & \quad $-1.46\times 10^{-3}$   & \quad   $-3.293\times 10^{-6}$    \\
\hline
$\lambda_5$     & \quad   $-4.177\times 10^{-3}$       & \quad  $-2.07\times 10^{-3}$  & \quad   $-3.191\times 10^{-3}$   \\
\hline
$m_{\eta}^{2}$ [GeV]      & \quad   $ 1.851\times 10^{5}$   & \quad $1.276\times 10^{5}$  & \quad   $ 1.234\times 10^{5}$    \\
\hline
$m_{\eta_R}$ [GeV]      & \quad   $430.141$   & \quad  $357.093$   & \quad   $351.087$   \\
\hline
$m_{\eta_I}$ [GeV]      & \quad   $430.435$   & \quad  $357.269$   & \quad   $351.362$   \\
\hline
$m_{\eta^\pm}$ [GeV]      & \quad   $430.288$   & \quad $357.243$  & \quad   $ 351.224 $  \\
\hline
$m_{N_1}$ [GeV]    & \quad   $434.197$    & \quad $357.175$    & \quad   $ 351.134$   \\
\hline
$c\tau_{\eta^{\mp}}$ [mm]    & \quad   $16.859$    & \quad  $14.587$   & \quad   $ 28.412$    \\
\hline
$\sigma(pp\rightarrow \eta \eta j)$ [fb]    & \quad  $2.525$   & \quad    $5.44$ & \quad   $5.81$     \\

\hline
$N = \sigma \times BR \times \mathcal{L}\times \epsilon$    & \quad  $19.392$   & \quad    $33.474$ & \quad   $77.811$     \\

\hline
$\Omega h^{2}$     & \quad   $0.121$   & \quad   $0.121$   & \quad   $0.119$   \\
\hline
\end{tabular}}
\caption{Relevant model spectrum and parameters for three representative benchmarks, B3 (partial relic), B4 and B5 (total relic), with $\eta^{\pm}$ as a long-lived particle. Cross-sections with $\eta=\eta^{\pm}, \eta_{R}$ at $pp$ were calculated at $\sqrt{s}= 14$ TeV with \texttt{MadGraph}~\cite{Alwall:2014hca}. Number of expected events at $\mathcal{L}=3000/$fb.}
\label{Table:etaBMs}
\end{table}


\section{Conclusions}
\label{sec:summary}

We revisit the scotogenic model with a thermal scalar dark matter candidate in a region of parameter space that has had less attention, for a dark matter mass $m_{\eta_{R}}$ below 500 GeV.  We find that a mass splitting of $\Delta m_{N_{1}} \equiv m_{N_{1}} - m_{\eta_{R}}$  below $\approx 20$ GeV and $\Delta m_{\eta^\pm} \equiv m_{\eta^{\pm}}  - m_{\eta_{R}}$ below $\approx 1$ GeV leads to regions in model parameter space where a correct DM relic abundance can be satisfied, while also satisfying current constraints from lepton flavour violation, neutrino physics, direct and indirect detection. Such small mass splittings naturally predicts long-lived particles.

Motivated by the possibility to distinguish the scotogenic model from other scenarios with similar scalar phenomenology (such as the Inert Higgs Doublet Model), we focus on regions where the mass splitting between $\eta^{\pm}$ and $N_{1}$, $\Delta m_{\eta N}$, is below $\sim 1$ GeV, as then the lightest fermion present in the scotogenic model, $N_{1}$, could have a macroscopic lifetime while maximizing the total relic abundance. We find such detection to be very unlikely due to extremely low rates coming from the smallness of the yukawa couplings between $N_{1}$ and the SM leptons.  On the other hand, if $\Delta m_{\eta^\pm} \equiv m_{\eta^{\pm}}  - m_{\eta_{R}}$ is $\sim \mathcal{O}(100)$ MeV, a long-lived charged scalar can have proper decay distances in the millimeter to centimeter range,  and could give rise to disappearing charged track signatures from its decay to dark matter and a soft pion, while still satisfying a DM candidate below 500 GeV. This scenario could be tested within the reach of the high-luminosity LHC.

\acknowledgments{We thank Sebastian Urrutia for discussions in the early stages of this work and Stefano Gariazzo for the discussion on dark matter standard cosmology. We are also very grateful to Valentina De Romeri for useful comments and the reading of this manuscript. I.M.A. acknowledges support by ANID-Chile FONDECYT Grant No. 3210145. G.C. acknowledges support by ANID-Chile FONDECYT Grant No. 11220237. G.C. and M.A.D also acknowledge support by ANID – Millennium Science Initiative Program ICN2019\_044.}




\bibliographystyle{apsrev4-1}
\bibliography{main}

\end{document}